# Generalized Secure Distributed Source Coding with Side Information[1]

Somayeh Salimi[1], Mahmoud Salmasizadeh[2], Mohammad Reza Aref[1]

[1]ISSL Lab., Dept. of Electrical Engineering, Sharif University of Technology, Tehran, Iran

[2]Electronics Research Center, Sharif University of Technology, Tehran, Iran

Email: salimi@ee.sharif.edu, salmasi@sharif.edu, aref@sharif.edu

**Abstract.** In this paper, new inner and outer bounds on the achievable compression-equivocation rate region for generalized secure data compression with side information are given that do not match in general. In this setup, two senders, Alice and Charlie intend to transmit information to Bob via channels with limited capacity so that he can reliably reconstruct their observations. The eavesdropper, Eve, has access to one of the channels at each instant and is interested in the source of the same channel at the time. Bob and Eve also have their own observations which are correlated with Alice's and Charlie's observations. In this model, two equivocation and compression rates are defined with respect to the sources of Alice and Charlie. Furthermore, different special cases are discussed where the inner and outer bounds match. Our model covers the previously obtained results as well.

## 1 Introduction

Recently, much theoretical research has been dedicated to the problems of nodes with dense distribution and resource constraints. In such networks, nodes compress correlated data separately, without collaboration, which is known as distributed source coding. The nature of sensitive data in these networks necessitates secure compression while meeting quality of service requirement.

[1] Part of this work was accepted at WCC2009 [12]

In the classical wiretap channel model, considered by Wyner in [1], nonzero secrecy rate can be achieved without using a secure key if the intended receiver has a better quality communication channel than the eavesdropper. The idea of generating a shared secret key from the correlated observations has been explored vastly so far. Maurer [2] and Ahlswede and Csiszar [3] were researchers in this subject. On the other hand, in some applications like sensor networks, it is needed that the correlated information from different sources can be reconstructed by a node which has some other source of information. Simultaneously, the information sources should be kept secret from an eavesdropper as much as possible.

In this paper, we explore the above mentioned problem, where two nodes must separately compress their sources in such a way that the eavesdropper with side information learns as little as possible about them. In our model, a general secure distributed compression problem is considered in which two transmitters Alice and Charlie, with correlated observations, intend to send information to a receiver, Bob, over noiseless channels with limited capacity in such a way that he can reconstruct both sources reliably. Also, there is an eavesdropper, Eve, who listens to either Alice or Charlie's channel, one at a time, and when she listens to each of the sender's channel, she is only interested in learning information about that sender's source. Bob and Eve have their own side information correlated to Alice's and Charlie's observations.

In the sequel, we aim at information source when we use the term source.

### 1.1 Related Works

In [4], Yamamoto addresses lossy compression with security constraints over a noisy broadcast channel in which the users share a secure key. It was shown that if the source is compressed at the first step and then encrypted using the secure key and finally transmitted over the noisy channel using wiretap channel code, the optimal strategy is selected. In some other works, the communication channels are considered noiseless. For example in [5] and [6], Yamamoto investigated the scheme where a sender observes the outcomes of two correlated sources and wishes to send information in such a way that one of the sources can be reconstructed at the receiver but the other is kept as secret as possible. A simplified but significant problem has been addressed by Prabhakaran and Ramchandran in [7] where Alice intends to send information to Bob to enable him to reconstruct her source and keep eavesdropper, Eve, as ignorant as possible about her source. In this problem, Bob and Eve have access to side information arbitrarily correlated to Alice's source, and the minimum leakage rate in secure lossless compression is explored. The significance of [7] is explaining the point that in the case of arbitrarily correlated side information at the eavesdropper, the usual Slepian-Wolf compression is not always sufficient.

Secure lossless compression of two correlated sources is investigated in [8], where the related information of each source is sent over one channel and the eavesdropper has access to only one of the channels stream at any instant and wishes to get information about the sender's source of that channel at that time. The problem explored in [8], by Luh and Kundur, is a simplified case in which eavesdropper has no side information and so Slepian-Wolf coding suffices to setup minimum leakage. In this situation, the compression-equivocation capacity region is given. In another scenario in [9] and [10], Gunduz et. al. have investigated a situation where Eve has only access to the Alice's channel, and the other channel is secure. In this case, equivocation is calculated with respect to Alice's source. In [9], transmitted data from Charlie acts as side information at Bob but in [10], there is an additional condition that Charlie's source can be reconstructed by Bob. In both cases, inner and outer bounds on achievable compression-equivocation are given in [9] and [10].

### 1.2 Our Contribution

It can be seen that the scenarios considered in [8], [9] and [10] can be regarded as special cases of a generalized scenario in which Alice's and Charlie's sources should be reconstructed by Bob. The information related to Alice's and Charlie's sources is sent over relevant noiseless channel with limited capacity. In this scenario, Bob and Eve have access to side information arbitrarily correlated with Alice's and Charlie's sources. Similar to [8], Eve has access to only one of the channels stream at any instant and wishes to get information about the sender's source of that channel at that time. So, in this situation, two equivocation rates are defined and the inner and outer bounds on achievable compression-equivocation rates $(R_A, R_C, \Delta_A, \Delta_C)$ are explored in which $R_A$ and $R_C$ are the capacity of Alice's and Charlie's channels and $\Delta_A$ and $\Delta_C$ are the equivocations of Eve with respect to Alice's and Charlie's sources, respectively. In this generalized scenario, the Slepian-Wolf coding alone is not optimal and must be combined with random coding. Also, different cases are discussed and it can be seen that our results contain the results of the above mentioned references as the special cases.

The paper is organized as follows. The generalized model is introduced in section 2. In section 3, inner and outer bounds of compression-equivocation rate region are given in theorems 1 and 2, respectively, which generalize the well known Slepian-Wolf region to include secrecy constraints. Different scenarios based on the availability of the side information at the nodes are considered in section 4. The conclusion and the proofs of the theorems are included in section 5 and appendix, respectively.

## 2 Generalized Model

In the model shown in Fig.1, it is assumed that Alice and Charlie have access to observations of the length-$N$ correlated source sequences $A^N$ and $C^N$, respectively, and intend to transmit required information to Bob via noiseless but finite capacity channels so that Bob can reliably reconstruct these sources with access to side information sequences $B^N$. The eavesdropper, Eve, with access to correlated side information $E^N$ can intercept only one of the channels at a time and at that time, is only interested in obtaining information about the related source of that channel. It is assumed that the observations $A^N, B^N, C^N$ and $E^N$ are generated independently and identically distributed (i.i.d.) with joint probability distribution $P_{ABCE}(a,b,c,e)$ over the finite alphabet $\mathcal{A}\times\mathcal{B}\times\mathcal{C}\times\mathcal{E}$. In this model, while Alice and Charlie want to transmit required information to Bob simultaneously, each of them attempts to minimize the rate of information learned about her/his source by Eve. Concerning this, the equivocation of Eve with respect to each sources can be defined when Eve has access to the related channel. Throughout the paper, we assume that all the transmissions are authenticated, i.e., the eavesdropper is passive.

***Definition 1***: A $(M_A, M_C, N)$ code for secure compression of the sources includes two stochastic encoding functions for Alice and Charlie, respectively, $f_A: \mathcal{A}^N \rightarrow I_{M_A}$ and $f_C: \mathcal{C}^N \rightarrow I_{M_C}$ as well as a decoding function $g: I_{M_A} \times I_{M_C} \times \mathcal{B}^N \rightarrow \mathcal{A}^N \times \mathcal{C}^N$ for Bob. In this setup, the equivocation of Eve with respect to Alice's and Charlie's sources is defined as $\frac{1}{N} H(A^N \mid f_A(A^N), E^N)$ and $\frac{1}{N} H(C^N \mid f_C(C^N), E^N)$, respectively, and the error probability is defined as $\Pr\{g(f_A(A^N), f_C(C^N)) \neq (A^N, C^N)\}$.

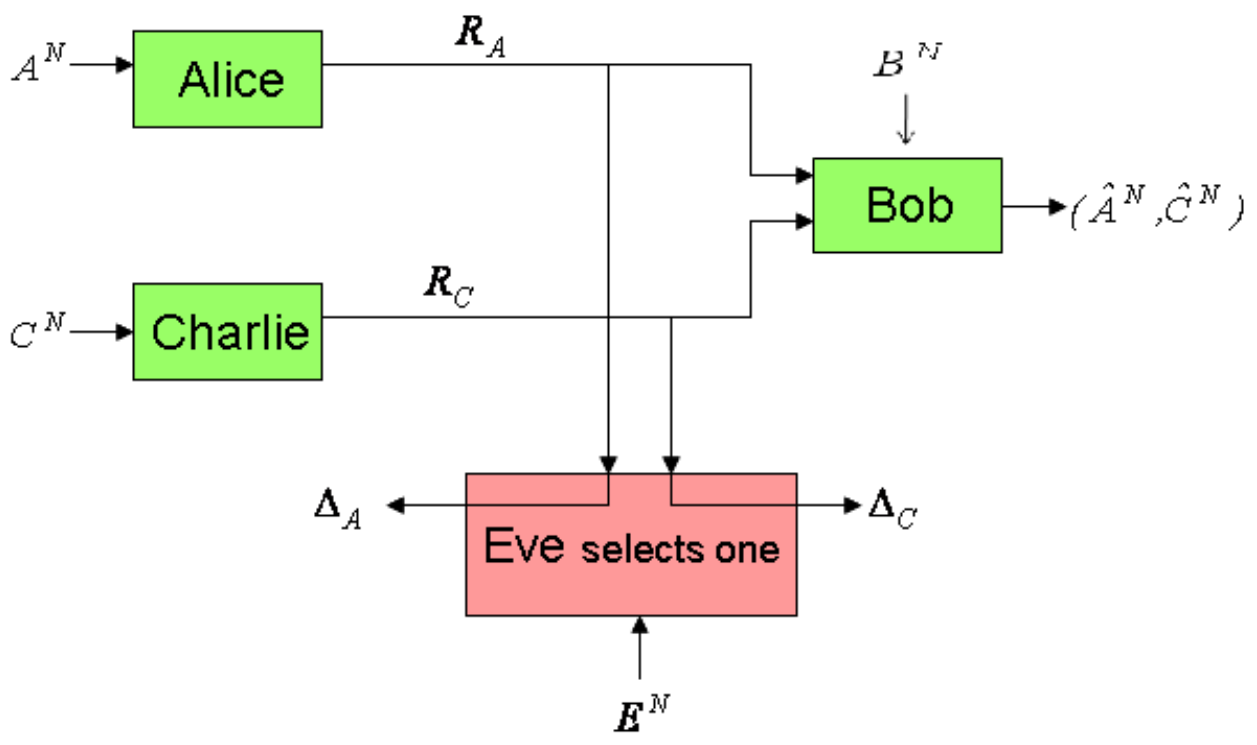

**Fig. 1.** Secure compression of two sources

***Definition 2***: A rate quadruple $(R_A, R_C, \Delta_A, \Delta_C)$ is said to be achievable if for any $\varepsilon > 0$, there exists a $(M_A, M_C, N)$ code such that:

$\log(M_A) \le N(R_A + \varepsilon)$,

$\log(M_C) \le N(R_C + \varepsilon)$,

$H(A^N \big| f_A(A^N), E^N) \ge N(\Delta_A - \varepsilon)$ and

$H(C^N \big| f_C(C^N), E^N) \ge N(\Delta_C - \varepsilon)$, where $P_e^N < \varepsilon$.

The closure of all achievable rate quadruples $(R_A, R_C, \Delta_A, \Delta_C)$ is compression-equivocation capacity region for which the inner and outer bounds are represented in section 3.

## 3 Generalized Secure Compression of Sources

If security is ignored in the problem described in the previous section, then it is reduced to a Slepian-Wolf coding of correlated sources [11]. Considering security in this problem makes it more complicated and it can not be considered as a simple extension of Slepian-Wolf problem. Some simpler cases of this problem are explored in [8],[9] and [10]. In [8], Eve has no side information and so, Slepian-Wolf coding strategy suffices. In [9], the situation is investigated where only Alice's channel is intercepted by Eve while there is no need for Charlie's source to be reconstructed by Bob, i.e. the information sent by Charlie has the role of side information at Bob and leads to less information leakage of Alice's source. In [10] the condition of reconstructing Charlie's source by Bob is added. In both last cases, Eve has access to side information.

For the setup described in definition 2, we give the inner and outer bounds of all achievable $(R_A, R_C, \Delta_A, \Delta_C)$ which do not match in general. In this case, the Slepian-Wolf coding strategy does not suffice and the coding strategy of [10] is followed with the difference that security is an important issue for both senders and the scenario of [10] should be modified appropriately.

***Theorem 1 (inner bound)***: In the described setup $(R_A, R_C, \Delta_A, \Delta_C)$ is achievable if:

$$R_A \geq H(A|V,B) - I(A;C|U,V,B) \quad (1)$$
$$R_C \geq H(C|U,B) - I(A;C|U,V,B) \quad (2)$$
$$R_A + R_C \geq H(A,C|B) \quad (3)$$
$$\Delta_A \leq \max\{I(A;B,C|U) - I(A;E|U)\} \quad (4)$$
$$\Delta_C \leq \max\{I(A,B;C|V) - I(C;E|V)\} \quad (5)$$
$$\Delta_A + \Delta_C \leq \max\{I(A,C;U,V,B) + I(A;C) - I(A;U,E) - I(C;V,E)\} \quad (6)$$
$$R_A + \Delta_A \geq H(A|E) \quad (7)$$
$$R_C + \Delta_C \geq H(C|E) \quad (8)$$

where the maximization is over the auxiliary random variables $U$ and $V$ that are according to the joint distribution $p(a,b,c,e,u,v) = p(a,b,c,e)p(u|a)p(v|c)$.

***Theorem 2 (outer bound)***: In the described setup, if $(R_A, R_C, \Delta_A, \Delta_C)$ is achievable, the equations (9)-(18) hold for some auxiliary random variables $U$ and $V$ that form Markov chains as $U - A - (B,C,E)$ and $V - C - (A,B,E)$.

$$R_A \geq H(A|V,B) - I(A;C|U,V,B) \quad (9)$$
$$R_C \geq H(C|U,B) - I(A;C|U,V,B) \quad (10)$$
$$R_A + R_C \geq H(A,C|B) \quad (11)$$
$$\Delta_A \leq \max\{I(A;B,V|U) - I(A;E|U)\} \quad (12)$$
$$\Delta_C \leq \max\{I(C;B,U|V) - I(C;E|V)\} \quad (13)$$
$$\Delta_A \leq R_C + I(A;B) - H(C|A,B) \quad (14)$$
$$\Delta_C \leq R_A + I(C;B) - H(A|B,C) \quad (15)$$
$$\Delta_A + \Delta_C \leq I(A;C) + I(A,C;B) \quad (16)$$
$$R_A + \Delta_A \geq H(A|E) \quad (17)$$
$$R_C + \Delta_C \geq H(C|E) \quad (18)$$

The detailed proofs of both theorems are given in the appendix but some discussion is followed.

In the mentioned problem, Alice and Charlie attempt to increase equivocation of Eve with respect to their own sources. In [10] Charlie's channel was secure and he used his channel capacity as much as possible so that Alice could keep her source as secret as possible from Eve. A significant difference of our problem with the problem in [10] is that Charlie is also attentive of his source security and so, most usage of his channel capacity is not a good strategy, necessarily. This makes the problem more complex and according to equations (6) and (16), there is a trade-off between the equivocation rates. In fact, in the achievability scheme, it can be assumed that first, Alice and Charlie transmit the auxiliary random variables $U$ and $V$, respectively, which have distributions $p(u|a)$ and $p(v|c)$. After that, they launch to transmit remainder of information which is required for Bob to reconstruct both sources. For this purpose, according to Slepian-Wolf theorem, Alice and Charlie should transmit some information with the overall rate $H(A,C|B,U,V)$ at least, which is sent via random binning. This rate is divided between Alice and Charlie and this division determines the trade-off according to equation (6).

Using the auxiliary random variables can potentially result in higher equivocation rates. This fact is shown in [7] via an example in the case that there is only one sender. In this binary erasure example, it is proved that transmitting part of information via an auxiliary random variable and the other part via Slepian-Wolf coding leaks less information to the eavesdropper compared to the situation where information is sent via Slepian-Wolf coding entirely. This fact is true in the case of two senders and two equivocation rates. However, when eavesdropper has no side information, Slepian-Wolf coding suffices and the maximum equivocation rates can be achieved by constant auxiliary random variables [7]. In addition, there are some other situations where constant random variables can result in maximum equivocation rates.

***Corollary 3.1***: When eavesdropper's side information $E$ is physically degraded with respect to $B$ i.e. if $A-B-E$ or $C-B-E$ form Markov chains, then $U$ or $V$ can be chosen constant, respectively. If both of Markov chains hold, then choosing both auxiliary random variables constant is optimum.

We prove this corollary for auxiliary random variable $U$. For the random variable $V$ the proof is similar. From equation (12) of theorem 2, the equivocation rate $\Delta_A$ will be upper bounded as:

$$\begin{aligned}
\Delta_A &\le \max\{I(A;B,V|U) - I(A;E|U)\} \le \max\{I(A;B,C|U) - I(A;E|U)\} \qquad (19)\\
&\le \max\{I(A;B,C|U,E)\} = \max\{H(B,C|U,E) - H(B,C|U,A,E)\}\\
&= \max\{H(B,C|U,E) - H(B,C|A,E)\} \qquad (20)\\
&\le H(B,C|E) - H(B,C|A,E) = I(A;B,C|E)
\end{aligned}$$

where equation (19) follows from the data processing inequality and equation (20) is the direct result of Markov chain $U-A-(B,C,E)$. On the other hand, this upper bound is achievable by setting $U=$ Constant and applying Markov chain $A-B-E$ in equation (4) of theorem 1.

## 4 Special Cases

Now, we investigate some special cases. The first and also the simplest case is when eavesdropper has no side information or in other words, $E$ is constant.

***Case 1.Eavesdropper with no side information***: In this situation, the following corollary can be deduced.

***Corollary 4.1***: When there is no side information at Eve, the inner and outer bounds match each other. The compression-equivocation capacity region is characterized by:

$$R_A \geq H(A|B,C) \quad (21)$$

$$R_C \geq H(C|A,B) \quad (22)$$

$$R_A + R_C \geq H(A,C|B) \quad (23)$$

$$[H(A)-R_A]^+ \leq \Delta_A \leq \min\{I(A;B,C), R_C + I(A;B) - H(C|A,B)\} \quad (24)$$

$$[H(C)-R_C]^+ \leq \Delta_C \leq \min\{I(A,B;C), R_A + I(C;B) - H(A|B,C)\} \quad (25)$$

$$\Delta_A + \Delta_C \leq I(A,C;B) + I(A;C) = I(A;B,C) + I(B;C) = I(A,B;C) + I(A;B) \quad (26)$$

The achievability and converse of the corollary can be considered as a special case of theorems 1and 2 by setting the auxiliary random variable $U$ and $V$ constant and using Slepian-Wolf binning. It can be seen that this special case coincides in [8, theorem 1].

Some other special cases can be assumed when side information of Bob or Eve or both is accessible at Alice or Charlie. Hence, three cases can be considered when Alice has access to Bob's or Eve's side information or both. For simplicity, in these three cases, it is assumed that the channels from Alice and Charlie to Bob have infinite capacity and only the equivocation rates are considered.

***Case 2.Eavesdropper's side information at Alice***: In this case, it is assumed that Eavesdropper's side information (source $E$ ) is available at Alice and the following corollary can be obtained.

***Corollary 4.2***: If Eve's side information is available at Alice, $(R_A, R_C, \Delta_A, \Delta_C)$ is achievable if and only if:

$$\Delta_A \leq I(A;B,C|E) \quad (27)$$

$$\Delta_C \leq I(C;A,B|E) \quad (28)$$

$$\Delta_A + \Delta_C \leq I(A;B,C|E) + I(B;C|E) = I(A,B;C|E) + I(A;B|E) \quad (29)$$

It can be shown that in this case, the best strategy is to establish $U = E$ and $V$ as constant value and transmit remainder information via Slepian-Wolf random binning. The rate of this remainder information is $H(A,C|B,E)$. This rate must be sent by Alice and Charlie and depending on the portion of transmitted information by each of them, there is a tradeoff between the equivocation rates.

Achievability of corollary 2 can be obtained by setting $U=E$ and $V$ as constant value in theorem 1. For the converse part of corollary 2, from equation (12) of theorem 2, we have:

$$\Delta_A \le \max\{I(A;B,V|U)-I(A;E|U)\} \le \max\{I(A;B,V|U,E)\} \le \max\{I(A;B,C|U,E)\} \le I(A;B,C|E) \quad (30)$$

Equation (30) is the result of $U-(A,E)-(B,C)$ Markov chain. We note that when Alice has access to Eve's side information, the source $A$ will be replaced with $A,E$. Similarly equations (28) and (29) can be deduced.

***Case 3.Bob's side information at Alice***: In this case, it is assumed that Bob's side information (source $B$) is available at Alice and the inner and outer bounds of theorems 1 and 2 hold where the auxiliary random variables $U$ and $V$ are according to the joint distribution $p(a,b,c,e,u,v)=p(a,b,c,e)p(u|a,b)p(v|c)$ in theorem 1 and form Markov chains as $U-(A,B)-(C,E)$ and $V-C-(A,B,E)$ in theorem 2.

It should be noted that the availability of either Eve's or Bob's side information at Alice enlarges the space of auxiliary random variable $U$ and can potentially result in higher equivocation rates at the eavesdropper. From the Slepian-Wolf source coding, it is known that the availability of receiver's side information at the senders does not result in compression rates improvement, but this is not true in the case of equivocation rates. We show via an example that availability of Bob's side information at Alice increases equivocation rate of Eve with respect to Alice's source. This example is an extension of the example in lemma 4.2 in [10].

It is evident that when $A$ is independent of $(B,C)$ and Alice does not have access to $B$, then $\Delta_A=0$. However when she has access to $B$, she can use it as a key to encrypt her message.

***Corollary 4.3***: When $A$ is independent of $(B,C)$, $B$ is independent of $E$ and Alice has access to $B$, then $(\Delta_A,\Delta_C)$ is achievable if and only if:

$$\Delta_A \le \min\{H(B)-I(A;E),H(A|E)\} \quad (31)$$
$$\Delta_C \le \max\{I(C;B|V)-I(C;E|V)\} \quad (32)$$

where the maximization is over the auxiliary random variable $V$ which forms Markov chain as $V-C-(B,E)$.

For achieving the equivocation rate of equation (31), Alice uses $B$ as secret key to encrypt her message. In fact, one time pad encryption system is established due to the fact that $B$ is independent of $E$. First, Alice compresses the source $A$ and then encrypts it by xoring with the bits of the key $B$. In this way, two situations can be occurred; if $H(B)\ge H(A)$, then Alice can encrypt her message entirely with the key and hence, equivocation rate at Eve is $H(A|E)$ with respect to Alice's source. If $H(B)<H(A)$ then, Alice should send some part of her message as cleartext and so, equivocation rate at Eve with respect to Alice's source is $H(A|E)-(H(A)-H(B))=H(B)-I(A;E)$. Achievability of equation (32) follows from equation (5) of theorem 1 and independency of $A$ and $C$. The converse can be obtained from equations (12) and (13) of theorem 2.

***Case 4.Bob's and Eve's side information at Alice***: In this case Bob's and Eve's side information (source $B$ and $E$) is available at Alice. Also in this case, the inner and outer bounds of theorems 1 and 2 hold with the difference that the auxiliary random variables $U$ and $V$ that are according to the joint distribution $p(a,b,c,e,u,v)=p(a,b,c,e)p(u|a,b,e)p(v|c)$ in theorem 1 and form Markov chains as $U-(A,B,E)-C$ and $V-C-(A,B,E)$ in theorem 2.

Similar to two previous cases, availability of Eve's and Bob's side information at Alice can potentially result in higher equivocation rates at the eavesdropper. However, if the observation of Eve is a physically degraded version of Bob's information, i.e. $A-B-E$ form a Markov chain, then providing Eve's side information to Alice can not improve the equivocation rates [9].

Another special case occurs when Eve's side information is available at Bob in addition to his source $B$.

***Case 5.Eve's side information at Bob***: In this case, availability of Eve's side information (source $E$) at Bob makes Eve's side information physically degraded with respect to Bob's side information because the trivial Markov chains $A-(B,E)-E$ and $C-(B,E)-E$ hold. Hence according to corollary 3.1, Slepian-Wolf coding suffices for maximizing equivocation rates.

***Corollary 4.4***: When Eve's side information is available at Bob, then the quadruple rate $(R_A,R_C,\Delta_A,\Delta_C)$ is achievable if and only if:

$$R_A \geq H(A|B,C,E) \tag{33}$$
$$R_C \geq H(C|A,B,E) \tag{34}$$
$$R_A+R_C \geq H(A,C|B,E) \tag{35}$$
$$0 \leq \Delta_A \leq I(A;B,C|E) \tag{36}$$
$$0 \leq \Delta_C \leq I(C;A,B|E) \tag{37}$$
$$0 \leq \Delta_A+\Delta_C \leq I(A;B,C|E)+I(B;C|E)=I(A,B;C|E)+I(A;B|E) \tag{38}$$
$$R_A+\Delta_A \geq H(A|E) \tag{39}$$
$$R_C+\Delta_C \geq H(C|E) \tag{40}$$

The achievability can be obtained from theorem 1 by replacing $B$ with $(B,E)$ and considering the auxiliary random variables $U$ and $V$ constant. The converse is followed by theorem 2.

## 5 Conclusion

In this paper, secure distributed compression of two sources was considered. In the studied model, the eavesdropper had access to side information and intercepted one of the channels at a time. In this model, two equivocation and compression rates were defined. The inner and outer bounds of rate quadruples $(R_A, R_C, \Delta_A, \Delta_C)$ were given in theorems 1 and 2 which did not generally match. This model contained the cases studied in [8], [9] and [10] and therefore could be referred to as a generalized model. In the coding scheme, combination of random coding and Slepian-Wolf were used and it was seen that there was a trade-off between two equivocation rates. Furthermore, some special cases are discussed in which the inner and outer bounds matched. The first special case happens when Eve has no side information. The compression-equivocation capacity region for this case in corollary 4.1 is an extension of the previously obtained one in [8]. Availability of Bob's or Eve's side information or both at Alice and availability of Eve's side information at Bob provide other special cases where compression-equivocation capacity region is given for each case. These cases were in agreement with the results obtained in this subject previously.

As the future work, the same problem can be regarded when the communication is two-way via public channel, i.e. Bob can also transmit data to Alice and Charlie. This leads in less leakage rate. In addition, the problem can be more generalized in a way that either there are multiple receiver with different side information that should decode the Alice's and Charlie's sources or there are multiple eavesdroppers with different side information that listen to one of the channels at a time.

## Appendix: Proofs

### Proof of Theorem 1[1]

In proof of achievability, we consider two rate quadruples $(R_A, R_C, \Delta_A, \Delta_C)$ and prove the equations of theorem 1 for them. We choose $U$ and $V$ with distributions $p(u|a)$ and $p(v|c)$ which satisfy the condition of theorem 1. Then, we generate $2^{N(I(A;U|B)+\varepsilon_1)}$ codewords of length $N$ , $U^N(w_1)$ for $w_1 \in \{1,...,2^{N(I(A;U|B)+\varepsilon_1)}\}$ with the distribution $\prod_{i=1}^{N} p(u_i)$ . Now, these codewords are randomly binned into $2^{N(I(A;U|V,B)+\varepsilon_2)}$ bins. The related bin index of a codeword $U^N(w_1)$ is denoted as $a(w_1)$. Also all $A^N$ sequences are randomly binned

[1] Like [10], we assume deterministic coding in the analysis for simplicity, but the proofs follow similarly for randomized coding which is modeled by assuming independent random variables at the terminals and deterministic coding functions that depend on these random variables.

into $2^{N(H(A|B,C,U)+\varepsilon_3)}$ bins and for a sequence $A^N$, the related bin index is denoted as $b(A^N)$. On the other hand, we similarly generate $2^{N(I(C;V|B)+\varepsilon_4)}$ codewords of length $N$, $V^N(w_2)$ for $w_2 \in \{1,...,2^{N(I(C;V|B)+\varepsilon_4)}\}$ with distribution $\prod_{i=1}^{N} p(v_i)$. Also all $C^N$ sequences are randomly binned into $2^{N(H(C|V,U,B)+\varepsilon_5)}$ and the bin index of a sequence $C^N$ is denoted as $c(C^N)$.

Now, we describe the encoding and decoding schemes for achievability of the following rate quadruple that satisfies equations (1), (3), (4), (6), (7) and (8) of theorem 1:

$$\begin{Bmatrix} R_A = H(A|V,B) - I(A;C|U,V,B) \\ R_C = H(C|U,B) + I(U;V|B) \\ \Delta_A = I(A;B,C|U) - I(A;E|U) \\ \Delta_C = I(C;U,B|V) - I(C;E|V) \end{Bmatrix}$$

in which $U$ and $V$ are random variables with the distribution of theorem 1.

For a typical observation of $A^N$, Alice finds a jointly typical sequence $U^N(w_1)$. It can be seen that this sequence is unique with a high probability. Then, for the sequences $U^N(w_1)$ and $A^N$, she detects the related index bins i.e. $a(w_1)$ and $b(A^N)$, respectively. These index bins are transmitted as Alice's encoding function and are received by Bob and Eve. For a typical observation of $C^N$, Charlie finds the related bin $c(C^N)$ and also the typical sequence $V^N(w_2)$ which is unique with high probability. Then he sends the index $(w_2)$ and $c(C^N)$ as his encoding function to Bob and Eve.

For decoding, Bob with access to side information sequences $B^N$ and index $(w_2)$, finds the related typical sequence $V^N(w_2)$ and then with this sequence, side information sequences $B^N$ and the bin index $a(w_1)$, he can find a typical $U^N(w_1)$ with high probability. With access to $U^N(w_1)$, side information sequences $B^N$ and the bin index $c(C^N)$, Bob decodes $C^N$ reliably with high probability and subsequently, he can correctly decode $A^N$ with access to the received bin index $b(A^N)$, the sequences $U^N(w_1)$, $C^N$ and $B^N$. The transmitted rates from Alice and Charlie are now calculated. If $\varepsilon_i \to 0$ for $i = 1,...,5$, we have:

$$R_A = I(A;U|V,B) + H(A|U,C,B) = H(A|V,B) - I(A;C|U,V,B) \quad (41)$$

$$\begin{aligned}R_C &= I(C;V|B) + H(C|U,V,B) = H(C|B) - I(C;U|V,B)\\ &= H(C|B) - H(U|V,B) + H(U|V,B,C)\\ &= H(C|B) - H(U|V,B) + H(U|B,C) \quad (42)\\ &= H(C|B) - H(U|B) + H(U|B,C) + I(U;V|B)\\ &= H(C|U,B) + I(U;V|B)\end{aligned}$$

$$\begin{aligned}R_A + R_C &= H(A|V,B) - I(A;C|U,V,B) + I(C;V|B) + H(C|U,V,B)\\ &= H(A|V,B) + H(C|A,V,B) + I(C;V|B)\\ &= H(A,C|V,B) + I(C;V|B) = H(C|B) - H(C|V,B) + H(A,C|V,B)\\ &= H(C|B) + H(A|V,B,C)\\ &= H(C|B) + H(A|C,B) = H(A,C|B)\end{aligned}$$

where equation (41) follows from the fact that $H(A|U,V,C,B) = H(A|U,C,B)$ which is the direct result of $V$ distribution. Also equation (42) exploits from the equality $H(U|V,B,C) = H(U|B,C)$.

Now, the equivocation rates can be lower bounded as:

$$\begin{aligned}N\Delta_A &= H(A^N|a(w_1), b(A^N), E^N)\\ &= H(A^N) - I(A^N; a(w_1), E^N) - I(A^N; b(A^N)|a(w_1), E^N)\\ &\geq H(A^N|a(w_1), E^N) - H(b(A^N))\\ &\geq H(A^N|U^N, E^N) - NH(A|U,B,C) - N\varepsilon_3 \quad (43)\\ &= N[H(A|U,E) - H(A|U,B,C) - \varepsilon_3] = N[I(A;B,C|U) - I(A;E|U) - \varepsilon_3] \quad (44)\end{aligned}$$

$$\begin{aligned}N\Delta_C &= H(C^N|w_2, c(C^N), E^N)\\ &= H(C^N) - I(C^N; w_2, E^N) - I(C^N; c(C^N)|w_2, E^N)\\ &\geq H(C^N|V(w_2), E^N) - H(c(C^N))\\ &\geq H(C^N|V^N, E^N) - NH(C|U,V,B) - N\varepsilon_5 = N[H(C|V,E) - H(C|U,V,B) - \varepsilon_5]\\ &= N[I(C;U,B|V) - I(C;E|V) - \varepsilon_5] \quad (45)\end{aligned}$$

Equation (43) follows from the data processing inequality. Also we have:

$$\begin{aligned}\Delta_A + \Delta_C &= H(A|U,E) - H(A|U,B,C) + H(C|V,E) - H(C|U,V,B) - (\varepsilon_3 + \varepsilon_5)\\ &= H(A|U,E) + H(C|V,E) - H(A,C|U,V,B) - (\varepsilon_3 + \varepsilon_5) \quad (46)\\ &= I(A,C;U,V,B) - I(A;U,E) - I(C;V,E) + I(A;C) - (\varepsilon_3 + \varepsilon_5)\end{aligned}$$

In deriving equation (46), the equality $H(A|U,B,C) = H(A|U,B,V,C)$ is used which is the consequence of $V$ distribution.

Finally, we have:

$$N\Delta_A = H(A^N \left| a(w_1), b(A^N), E^N \right.) = H(A^N \left| E^N \right.) - I(A^N; a(w_1), b(A^N) \left| E^N \right.)$$
$$\geq NH(A|E) - H(a(w_1), b(A^N)) \geq NH(A|E) - NR_A \qquad (47)$$
$$N\Delta_C = H(C^N \left| (w_2), c(C^N), E^N \right.) = H(C^N \left| E^N \right.) - I(C^N; w_2, c(C^N) \left| E^N \right.)$$
$$\geq NH(C|E) - H((V^N), c(C^N)) \geq NH(C|E) - NR_C \qquad (48)$$

Considering equations (41), (44), (45), (46), (47) and (48), equations (1), (3), (4), (6), (7) and (8) of theorem 1 are satisfied.

On the other hand, by symmetry, we describe the encoding and decoding schemes for achievability of the following rate quadruple that satisfies equations (2), (3), (5), (6), (7) and (8) of theorem 1:

$$\begin{cases} R_A = H(A|V,B) + I(U;V|B) \\ R_C = H(C|U,B) - I(A;C|U,V,B) \\ \Delta_A = I(A;V,B|U) - I(A;E|U) \\ \Delta_C = I(C;A,B|V) - I(C;E|V) \end{cases}$$

in which $U$ and $V$ are random variables with the distribution of theorem 1.

This coding and decoding is a bit modified version of the previous one. First, with fix distributions $p(u|a)$ and $p(v|c)$, the codewords $U^N(w_1)$ and $V^N(w_2)$ are generated similar to previous coding scheme. Then, all $A^N$ sequences are randomly binned into $2^{N(H(A|U,V,B)+\varepsilon_3)}$ bins in which for a sequence $A^N$, the related bin index is denoted as $b(A^N)$. Also $N$-length codewords $V^N(w_2)$ are randomly binned into $2^{NI(C;V|U,B)}$ bins and the related bin index of a codeword $V^N(w_2)$ is denoted as $a(w_2)$ and all $C^N$ sequences are randomly binned into $2^{NH(C|V,A,B)}$ and the related bin index is denoted as $c(C^N)$. For a typical observation of $A^N$, Alice finds a jointly typical sequence $U^N(w_1)$ which is unique with a high probability. Then, for the $U^N(w_1)$ and $A^N$, she detects the related index bins $b(A^N)$ and sends $(w_1)$ and $b(A^N)$ to Bob and Eve. For a typical observation of $C^N$, Charlie finds the typical sequence $V^N(w_2)$ and then finds the bin indices of these two sequence i.e. $a(w_2)$ and $c(C^N)$, respectively.

It can be seen that similar to the previous decoding scheme, Bob can reliably decode $A^N$ and $C^N$ with high probability. Now, the same parameters can be calculated (the procedure is the same as before and so details are omitted):

$$R_A = I(A;U|B) + H(A|U,V,B) = H(A|V,B) + I(U;V|B) \quad (49)$$
$$R_C = I(C;V|U,B) + H(C|V,A,B) = H(C|U,B) - I(A;C|U,V,B) \quad (50)$$
$$R_A + R_C = H(A,C|B) \quad (51)$$
$$N\Delta_A = H(A^N|w_1, b(A^N), E^N) = N[H(A|U,E) - H(A|U,V,B) - \varepsilon_3]$$
$$= N[I(A;V,B|U) - I(A;E|U) - \varepsilon_3] \quad (52)$$
$$N\Delta_C = H(C^N|a(w2), c(C^N), E^N) = N[H(C|V,E) - H(C|A,B,V) - \varepsilon_5]$$
$$= N[I(C;A,B|V) - I(C;E|V) - \varepsilon_5] \quad (53)$$
$$\Delta_A + \Delta_C = H(A|U,E) - H(A|U,V,B) + H(C|V,E) - H(C|V,A,B) - (\varepsilon_3 + \varepsilon_5)$$
$$= H(A|U,E) + H(C|V,E) - H(A,C|U,V,B) - (\varepsilon_3 + \varepsilon_5)$$
$$= I(A,C;U,V,B) - I(A;U,E) - I(C;V,E) + I(A;C) - (\varepsilon_3 + \varepsilon_5) \quad (54)$$
$$N\Delta_A = H(A^N|w_1, b(A^N), E^N) = H(A^N|E^N) - I(A^N; w_1, b(A^N)|E^N)$$
$$\geq NH(A|E) - H(w_1, b(A^N)) \geq NH(A|E) - NR_A \quad (55)$$
$$N\Delta_C = H(C^N|a(w_2), c(C^N), E^N) = H(C^N|E^N) - I(C^N; a(w_2), c(C^N)|E^N)$$
$$\geq NH(C|E) - H((a(w_2), c(C^N)) \geq NH(C|E) - NR_C \quad (56)$$

It can be seen that with this coding scheme, equations (2), (3), (5), (6), (7) and (8) of theorem 1 are satisfied with considering equations (50), (51), (53), (54), (55) and (56).

In both above coding schemes, the sum rates of compression rates and equivocation rates are the same.

Now, achievability of the two rate quadruples is proved. We avoid detailed proof of achievability of the total region of theorem 1and content ourselves with a proof scheme.

In both of the above coding scheme, at the first step, Alice and Charlie use auxiliary random variables $U$ and $V$ and send information with the total rate $I(A,C;U,V|B) = H(U,V|B) - H(U|A) - H(V|C)$. At the second step, Alice and Charlie should send total rate $H(A,C|U,V,B)$ so that Bob can reconstruct both of the sources. The portion of transmitted rates by each of Alice or Charlie at both steps determines the trade-off between the compression rates i.e. $R_A$ and $R_C$, while sum of them is fixed. Division of transmitted rates at the second step determines the trade-off between two equivocation rates i.e. $\Delta_A$ and $\Delta_C$, while sum of them is fixed. Therefore, four cases can happen that in all of them, the compression sum rate and the equivocation sum rate are fixed and equal to equations (3) and (6), respectively. We discus these cases as follows:

1-Alice sends information with the rate of $I(A;U|B) = H(A|B) - H(A|U,B)$ at the first step and then information with the rate of $H(A|U,V,B)$ at the second step. Charlie sends information with the rate of

$I(C;V|U,B) = H(C|U,B) - H(C|U,V,B)$ at the first step and then information with the rate of $H(C|V,A,B)$ at the second step. So we have:

$$\begin{cases} R_A = H(A|B) - I(A;V|U,B) = H(A|V,B) + I(U;V|B) \\ R_C = H(C|U,B) - I(A;C|U,V,B) \\ \Delta_A = I(A;V,B|U) - I(A;E|U) \\ \Delta_C = I(C;A,B|V) - I(C;E|V) \end{cases}$$

2-Alice sends information with the rate of $I(A;U|V,B) = H(A|V,B) - H(A|U,V,B)$ at the first step and then information with the rate of $H(A|U,V,B)$ at the second step. Charlie sends information with rate of $I(C;V|B) = H(C|B) - H(C|V,B)$ at the first step and then information with the rate of $H(C|V,A,B)$ at the second step. So we have:

$$\begin{cases} R_A = H(A|V,B) \\ R_C = H(C|B) - I(A;C|V,B) \\ \Delta_A = I(A;V,B|U) - I(A;E|U) \\ \Delta_C = I(C;A,B|V) - I(C;E|V) \end{cases}$$

3-Alice sends information with the rate of $I(A;U|B) = H(A|B) - H(A|U,B)$ at the first step and then information with the rate of $H(A|U,B,C)$ at the second step. Charlie sends information with the rate of $I(C;V|U,B) = H(C|U,B) - H(C|U,V,B)$ at the first step and then information with the rate of $H(C|U,V,B)$ at the second step. So we have:

$$\begin{cases} R_A = H(A|B) - I(A;C|U,B) \\ R_C = H(C|U,B) \\ \Delta_A = I(A;B,C|U) - I(A;E|U) \\ \Delta_C = I(C;U,B|V) - I(C;E|V) \end{cases}$$

4-Alice sends information with the rate of $I(A;U|V,B) = H(A|V,B) - H(A|U,V,B)$ at the first step and then information with the rate of $H(A|U,B,C)$ at the second step. Charlie sends information with the rate of $I(C;V|B) = H(C|B) - H(C|V,B)$ at the first step and then information with the rate of $H(C|U,V,B)$ at the second step. So we have:

$$\begin{cases} R_A = H(A|V,B) - I(A;C|U,V,B) \\ R_C = H(C|B) - I(C;U|V,B) = H(C|U,B) + I(U;V|B) \\ \Delta_A = I(A;B,C|U) - I(A;E|U) \\ \Delta_C = I(C;U,B|V) - I(C;E|V) \end{cases}$$

It can be seen that in the first and second cases the individual equivocation rates are the same but the transmitted rates are different. Using time sharing according to theorem g. of [11], the other rate quadruples between these points are achievable. The same is true for third and forth cases and hence, the achievability of the theorem 1 region is satisfied.

**Proof of theorem 2**

The stochastic functions $F_A$ and $F_C$ are defined as $F_A = f_A(A^N)$, $F_C = f_C(C^N)$. Fano's inequality leads to

$$H(A^N, C^N \mid B^N, F_A, F_C) \leq N\alpha(P_e^N) \tag{57}$$

where $\alpha(.)$ is a non-negative function with $\lim_{x\to 0}\alpha(x) = 0$. Now, we define:

$$U_i = (F_A, A_{i\text{-}1}, E_{i\text{-}1}) \tag{58}$$

$$V_i = (F_C, C_{i\text{-}1}, E_{i\text{-}1}) \tag{59}$$

It can be seen that $U_i - A_i - (B_i, C_i, E_i)$ and $V_i - C_i - (A_i, B_i, E_i)$ are Markov chains. We have:

$$\begin{aligned}
NR_A &\geq H(F_A) \geq H(F_A \mid B^N, F_C) = H(A^N, F_A \mid B^N, F_C) - H(A^N \mid B^N, F_A, F_C) \\
&\geq H(A^N \mid B^N, F_C) - N\alpha(P_e^N) \qquad (60) \\
&\geq \sum_{i=1}^{N} H(A_i \mid B^N, F_C, A^{i\text{-}1}, C^{i\text{-}1}) - N\alpha(P_e^N) \\
&= \sum_{i=1}^{N} H(A_i \mid B_i, F_C, A^{i\text{-}1}, C^{i\text{-}1}) - N\alpha(P_e^N) \qquad (61) \\
&= \sum_{i=1}^{N} H(A_i \mid B_i, F_C, C^{i\text{-}1}) - N\alpha(P_e^N) \qquad (62) \\
&\geq \sum_{i=1}^{N} H(A_i \mid B_i, F_C, C^{i\text{-}1}, E^{i\text{-}1}) - N\alpha(P_e^N) = \sum_{i=1}^{N} H(A_i \mid V_i, B_i) - N\alpha(P_e^N) \\
&\geq \sum_{i=1}^{N} H(A_i \mid V_i, B_i) - \sum_{i=1}^{N} I(A_i; C_i \mid U_i, V_i, B_i) - N\alpha(P_e^N)
\end{aligned}$$

where equation (60) follows from Fano's inequality and equation (61) from the memoryless property. Equation (62) is the result of Markov chain $A_i - (B_i, F_C, C^{i\text{-}1}) - A^{i\text{-}1}$.

Similarly, it can be deduced that:

$$NR_C \geq \sum_{i=1}^{N} H(C_i \mid U_i, B_i) - \sum_{i=1}^{N} I(A_i; C_i \mid U_i, V_i, B_i) - N\alpha(P_e^N) \tag{63}$$

Also we have:

$$NR_A + NR_C \ge H(F_A) + H(F_C) \ge H(F_A, F_C) \ge H(A^N, C^N \middle| B^N) - N\alpha(P_e^N) \qquad (64)$$

For the equivocation rates:

$$\begin{aligned}
N\Delta_A &= H(A^N \middle| F_A, E^N) = H(A^N \middle| F_A) - I(A^N; E^N \middle| F_A) \\
&= H(A^N \middle| F_A, F_C, B^N) + I(A^N; F_C, B^N \middle| F_A) - I(A^N; E^N \middle| F_A) \\
&\le N\alpha(P_e^N) + \sum_{i=1}^{N} I(A_i; F_C, B^N \middle| F_A, A^{i-1}) - \sum_{i=1}^{N} H(E_i \middle| F_A, E^{i-1}) + H(E^N \middle| A^N, F_A) \qquad (65) \\
&\le N\alpha(P_e^N) + \sum_{i=1}^{N} I(A_i; F_C, B^N \middle| F_A, A^{i-1}, E^{i-1}) - \sum_{i=1}^{N} H(E_i \middle| F_A, E^{i-1}, A^{i-1}) + H(E^N \middle| A^N) \qquad (66) \\
&\le N\alpha(P_e^N) + \sum_{i=1}^{N} I(A_i; F_C, B^N, C^{i-1} \middle| F_A, A^{i-1}, E^{i-1}) - \sum_{i=1}^{N} H(E_i \middle| F_A, E^{i-1}, A^{i-1}) + H(E^N \middle| A^N) \\
&= N\alpha(P_e^N) + \sum_{i=1}^{N} I(A_i; F_C, B^N, C^{i-1}, E^{i-1} \middle| F_A, A^{i-1}, E^{i-1}) - \sum_{i=1}^{N} H(E_i \middle| F_A, E^{i-1}, A^{i-1}) + H(E^N \middle| A^N) \\
&= N\alpha(P_e^N) + \sum_{i=1}^{N} I(A_i; F_C, B_i, C^{i-1}, E^{i-1} \middle| F_A, A^{i-1}, E^{i-1}) - \sum_{i=1}^{N} H(E_i \middle| F_A, E^{i-1}, A^{i-1}) + H(E^N \middle| A^N) \qquad (67) \\
&= N\alpha(P_e^N) + \sum_{i=1}^{N} I(A_i; B_i, V_i \middle| U_i) - \sum_{i=1}^{N} [H(E_i \middle| U_i) - H(E_i \middle| A_i)] \\
&= N\alpha(P_e^N) + \sum_{i=1}^{N} I(A_i; B_i, V_i \middle| U_i) - \sum_{i=1}^{N} I(A_i; E_i \middle| U_i) \qquad (68)
\end{aligned}$$

where (65) follows from Fano's inequality; (66) from memoryless property and the fact that conditioning reduces entropy. Equation (67) is derived from memoryless property and equation (68) from $U_i - A_i - E_i$ Markov chain.

In a similar way it can be shown that:

$$N\Delta_C \le N\alpha(P_e^N) + \sum_{i=1}^{N} I(C_i; B_i, U_i \middle| V_i) - \sum_{i=1}^{N} I(C_i; E_i \middle| V_i) \qquad (69)$$

On the other hand, we have:

$$\begin{aligned}
&H(A^N, C^N \middle| F_A, F_C, B^N) \le N\alpha(P_e^N) \\
&H(A^N, C^N \middle| F_A, F_C) - I(A^N, C^N; B^N \middle| F_A, F_C) \le N\alpha(P_e^N) \\
&H(F_A, F_C \middle| A^N, C^N) + H(A^N, C^N) - H(F_A, F_C) + I(A^N, C^N; B^N) \le N\alpha(P_e^N) \\
&H(A^N, C^N) - H(F_A) - H(F_C) + I(A^N, C^N; B^N) \le N\alpha(P_e^N) \\
&H(A^N) - H(F_A) - H(F_C) + H(C^N \mid A^N) + I(A^N, C^N; B^N) \le N\alpha(P_e^N) \\
&H(A^N \middle| F_A) - H(F_A \middle| A^N) - H(F_C) + H(C^N \middle| A^N) + I(A^N, C^N; B^N) \le N\alpha(P_e^N) \\
&H(A^N \middle| F_A, E^N) + H(C^N \middle| A^N) - NR_C + I(A^N, C^N; B^N) \le N\alpha(P_e^N) \\
&\Delta_A \le R_C + H(A) - H(A, C \middle| B) + \alpha(P_e^N) = R_C + I(A; B) - H(C \middle| A, B) + \alpha(P_e^N) \qquad (70)
\end{aligned}$$

Similarly, it can be seen that:

$$\Delta_C \le R_A + H(C) - H(A,C|B) + \alpha(P_e^N) = R_A + I(C;B) - H(A|B,C) + \alpha(P_e^N) \qquad (71)$$

Also we have

$$H(A^N, C^N | F_A, F_C, B^N) \le N\alpha(P_e^N)$$
$$H(A^N, C^N | F_A, F_C) - I(A^N, C^N; B^N | F_A, F_C) \le N\alpha(P_e^N)$$
$$H(F_A, F_C | A^N, C^N) + H(A^N, C^N) - H(F_A, F_C) - I(A^N, C^N; B^N) \le N\alpha(P_e^N)$$
$$H(A^N) + H(C^N) - I(A^N, C^N) - H(F_A) - H(F_C) - I(A^N, C^N; B^N) \le N\alpha(P_e^N)$$
$$H(A^N | F_A) - H(F_A | A^N) + H(C^N | F_C) - H(F_C | C^N) \le I(A^N; C^N) + I(A^N, C^N; B^N) + N\alpha(P_e^N)$$
$$H(A^N | F_A) + H(C^N | F_C) \le I(A^N; C^N) + I(A^N, C^N; B^N) + N\alpha(P_e^N)$$
$$H(A^N | F_A, E^N) + H(C^N | F_C, E^N) \le I(A^N; C^N) + I(A^N, C^N; B^N) + N\alpha(P_e^N)$$
$$\Delta_A + \Delta_C \le I(A;C) + I(A,C;B) + \alpha(P_e^N) \qquad (72)$$

Also, it can be deduced:

$$H(A|E) \le \frac{1}{N} H(A^N, F_A | E^N) = \frac{1}{N}[H(F_A | E^N) + H(A^N | E^N, F_A)]$$
$$\le \frac{H(F_A)}{N} + \Delta_A = R_A + \Delta_A \qquad (73)$$
$$H(C|E) \le \frac{1}{N} H(C^N, F_C | E^N) = \frac{1}{N}[H(F_C | E^N) + H(C^N | E^N, F_C)]$$
$$\le \frac{H(F_C)}{N} + \Delta_C = R_C + \Delta_C \qquad (74)$$

Finally, if a random variable $Q$ with uniform distribution over $\{1,2,...,N\}$ and other random variables as:

$$A = A_Q, B = B_Q, C = C_Q, E = E_Q, U = (U_Q, Q), V = (V_Q, Q)$$

are definedand, then the equations (9)-(18) are satisfied if we consider equations (62), (63), (64), (68), (69), (70), (71), (72), (73), (74) and let $N \to \infty$ and $P_e^N \to 0$ .